# New Insights on Stacking Fault Behavior in Twin Induced Plasticity from Meta-Atom Molecular Dynamics Simulations

Peng Wang[1], Shaofeng Xu[1], Jiabin Liu[2], Xiaoyan Li[3], Yujie Wei[4], Hongtao Wang[1]*,

Huajian Gao[5]*, Wei Yang[1]*

1 Institute of Applied Mechanics, Zhejiang University, Hangzhou 310027, China
2 College of Materials Science and Engineering, Zhejiang University, Hangzhou 310027, China
3 Department of Engineering Mechanics, Tsinghua University, Beijing 100084, China
4 LNM, Institute of Mechanics, Chinese Academy of Sciences, Beijing 100190, China
5 Division of Engineering, Brown University, Providence, RI 02912, USA

**ABSTRACT**

There is growing interest in promoting deformation twinning for plasticity in advanced materials, as highly organized twin boundaries are beneficial to better strength-ductility combination in contrast to disordered grain boundaries. Twinning deformation typically involves the kinetics of stacking faults, its interaction with dislocations, and dislocation - twin boundary interactions. While the latter has been intensively investigated, the dynamics of stacking faults has been less known. In this work, we report several new insights on the stacking fault behavior in twin induced plasticity from our meta-atom molecular dynamics simulation: The stacking fault interactions are dominated by dislocation reactions taking place spontaneously, different from the proposed mechanism in literatures; The competition among generating a single stacking fault, a twinning partial and a trailing partial dislocation is dependent on a unique parameter, *i.e.* stacking fault energy, which in turn determines deformation twinning behaviors. The complex twin-slip and twin-dislocation interactions demonstrate the dual role of deformation twins as both dislocation barrier and storage, potentially contributing to the high strength and ductility of advanced materials like TWIP steels where deformation twinning dominated plasticity accounts for the superb strength-ductility combination.

## I. INTRODUCTION

Understanding the deformation mechanisms of crystalline materials is not only scientifically interesting but also technologically important in developing materials with novel mechanical properties [1,2]. In this field, simplified material "models" that capture the essential features of real materials have always played a critical role. In 1940s, Bragg and Nye developed a bubble raft model which, for the first time, unveiled the "atomic" arrangements of dislocations and grain boundaries (GB) in polycrystalline materials [3]. In spite of great simplifications in several aspects, the bubble raft model has successfully captured dynamic behaviors of dislocation nucleation and GB migration during plastic deformation [4,5]. The exponentially increasing computing power over the last few decades has now made it possible to perform massively parallel molecular-dynamics (MD) simulations based on many-body semi-empirical interatomic potentials to capture more realistic and material-specific deformation behaviors. In particular, the embedded atom method (EAM) [6,7] and Finnis-Sinclair potentials (EAM-FS) [8], which account for local electron density and pairwise interactions, have been widely successful in modelling the deformation behaviors of nanocrystalline metals [9-14]. For example, MD simulations have led to a deformation mechanism map for representative single-element, nanocrystalline, face-centered-cubic (FCC) metals in terms of the average grain size and external stress [13]. However, with rapidly growing interests in multi-element and multi-component complex metallic alloys, including the so-called high entropy alloys[15,16], the development of interatomic potentials has now lagged far behind the computational power. This problem arises partly due to the intrinsic difficulty of the existing methods in constructing pairwise interaction terms in the interatomic potential, which becomes nearly intractable even for ternary alloy systems.

In contrast, the Ashby-type deformation mechanism maps have been widely adopted for guiding material designs and applications since 1970s [17]. The basic principle of such representations is to use mechanism-based constitutive equations to divide the deformation-parameter space into regions within which a single mechanism becomes rate controlling [13]. The deformation-parameter space usually consists of two of the three normalized parameters: stress, temperature and grain size. The underlying idea is that macroscopic mechanical responses often rely on only a few key parameters, such as the lattice constant, surface energy, stable and unstable stacking fault energies, elastic moduli, sublimation energy, vacancy formation energy, as well as the temperature and microstructure. Most of the very specific atomic details can be neglected so that the physical processes during alloy deformation can eventually be depicted by simplified models with essentially the same collective behaviors. With this principle in mind, here we propose an EAM-based meta-atom method that enables MD simulations of the mechanical behaviors of metallic alloys with simplified interatomic potentials. By this means, massively parallel atomistic simulations of three-dimensional polycrystals are performed to investigate the deformation mechanisms in twinning-induced-plasticity (TWIP) steel, a practically important alloy system with broad applications. The prevalence of deformation twinning in TWIP steel leads to both high strength and exceptional formability for industrial applications. We will show that the meta-atom molecular dynamics (MAMD) simulations is capable of capturing the microstructures and the typical deformation behaviors of TWIP steel as observed in experiments, as well as probing the parametric dependence of deformation mechanisms, with potential applications in guiding the alloy design with specific mechanical properties.

## II. META-ATOM METHOD

The proposed meta-atom method is based on the following conjecture: The mechanical properties of an alloy system are primarily governed by a finite set of material constants, such as the lattice constants, surface energy, stable and unstable stacking fault energies (SFE), elastic moduli, sublimation energy and vacancy formation energy. Once the completeness of this set of material constants is established, two systems with the same material constants should exhibit identical experimentally observable mechanical behaviors. In this way, a detailed distinction among various atomic species is discarded and an alloy system is represented by a set of meta-atoms with a single interatomic potential to fit all related material constants. The essence of this homogenization methodology lies with the assumption that there exists a length scale at which the behaviors of two systems are governed by the same set of measurable materials properties, in spite of atomic scale differences and inhomogeneities. A detailed discussion of the applicable size regime is provided in the online supporting materials.

Based on the above assumptions, the atomic interaction among meta-atoms is formulated based on the EAM-FS potential [8,18]:

$$U_i = F(\rho_i) + \frac{1}{2}\sum_{j \neq i} \phi(r_{ij}) \quad (1)$$

where $F(\rho_i)$ is the embedding energy of the $i$-th meta-atom, $\phi(r_{ij})$ the pair potential between meta-atoms $i$ and $j$, and $\rho_i$ the local electron density provided by the remaining atoms in the system. More details about the EAM potential can be found in the excellent review by Daw *et al.* [19]. The universality of the EAM potential provides adequate flexibility in constructing model materials with properties varying in a sufficiently large range, covering spectra of different classes of metallic alloys.

The open source MD code of Large-scale Atomic/Molecular Massively Parallel Simulator (LAMMPS)[20] was used for the simulations of fully three-dimensional polycrystal samples. The samples were constructed using the Voronoi algorithm with 27 random oriented grains. The samples with an average grain size of 20 nm have dimensions of $60 \times 60 \times 60$ nm$^3$, containing 18,200,000 atoms. Periodic boundary conditions were imposed in all three directions. All simulations were performed at the temperature of 300 K, with a fixed time step of 1 fs and a Nosé-Hoover thermostat. Before the uniaxial loading, the samples were fully relaxed for 100 ps. During loading, a 60% strain was applied at a constant strain rate of $2 \times 10^8$ s$^{-1}$.

## III. RESULTS AND DISCUSSION

Figures 1*A-C* plot the meta atom potential for a TWIP steel with composition of Fe-22w.t.%Mn [21,22]. Table 1 lists the material properties. To model crystal plasticity, the embedding function and the pair potential are optimized to fit the following parameters: lattice constants, elastic moduli ($C_{11}$, $C_{12}$ and $C_{44}$ for crystal with cubic symmetry), SFE ($\gamma_{sf}$), unstable SFE ($\gamma_{usf}$), low index surface energies ($\gamma_{111}$, $\gamma_{110}$, $\gamma_{100}$), sublimation energy ($E_{sub}$) and single vacancy formation energy ($E_{vac}$). The detailed procedure and the coefficients for the meta-atom potential are provided in the online supporting material.

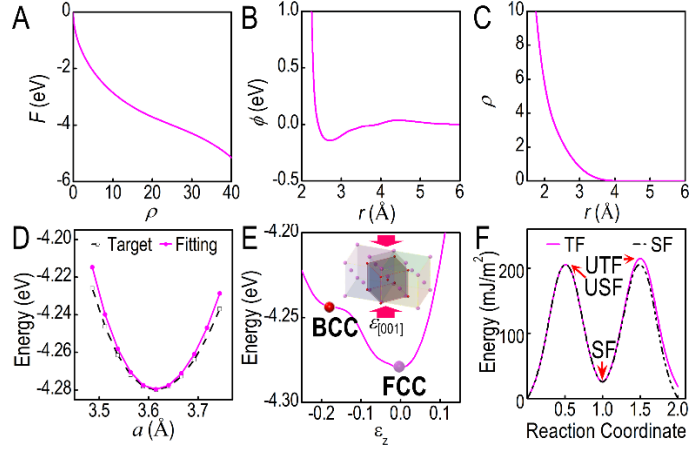

**FIG. 1.** The meta atom potential for a TWIP steel with composition of Fe-22w.t.%Mn. (*A*) The embedding functions $F(\rho)$, (*B*) the pair interaction functions $\phi(r)$ and (*C*) the electron density function $\rho(r)$. (*D*) Fitting to Rose's equation. (*E*) The energy landscape of Bain's transformation (inset) that connects FCC and body-centered-cubic configurations. (*F*) The corresponding generalized planar fault (GPF) energy for the stacking and twin fault planar defects. The GPF curve for an SF is obtained by rigidly shearing a perfect crystal along <112> on a (111) plane, and the GPF curve for a twin fault (TF) is obtained by rigidly shearing a perfect crystal containing a pre-existing SF.

**Table 1**. Fitting results of the meta atom potential for TWIP steel.

| Material property | Target value | Meta-atom TWIP steel |
|---|---|---|
| $a$ (fcc) (Å) | 3.615 [23] | 3.614 |
| $E_{coh}$ (eV/atom) | -4.28 [24] | -4.28 |
| $C_{11}$ (GPa) | 175 [23] | 179 |
| $C_{12}$ (GPa) | 83 [23] | 101 |
| $C_{44}$ (GPa) | 97 [23] | 108 |
| $E_{vac}$ (eV) | 1.7 [25] | 1.69 |
| $\gamma_{111}$ (mJ/m$^2$) | 1900 [26] | 927 |
| $\gamma_{110}$ (mJ/m$^2$) | 2100 [26] | 1308 |
| $\gamma_{100}$ (mJ/m$^2$) | 2000 [26] | 1054 |
| $\gamma_{sf}$ (mJ/m$^2$) | 19 [27] | 19 |
| $\gamma_{usf}$ (mJ/m$^2$) | 200 [28] | 211 |

The anharmonicity of a crystal is guaranteed if the cohesive energy as a function of

lattice constant satisfies Rose's equation, the so-called universal state equation of metals (Fig. 1D) [29]:

$$E(r_{WS}) = E_{sub} \cdot (-1 - a^* - 0.05 a^{*3}) \cdot e^{-a^*} \qquad (2)$$

where $a^* \equiv (r_{WS} - r_{WSE})/l$ and $l \equiv \sqrt{\dfrac{E_{sub}}{12\pi B r_{WSE}}}$, $r_{WS}$, $r_{WSE}$ and $B$ denoting, respectively, the radius of the Wigner-Seitz sphere containing an average volume per atom, the equilibrium Wigner-Seitz radius and the isothermal bulk modulus. The route of Bain's transformation [30-32], accounting for structural transformation with minimal atomic motion, is also considered in the optimization scheme in order to retain the stability of the lattice structure (Fig. 1E). The difference in free energies and the phase transformation barrier are both adopted as fitting parameters in order to ensure structural stability. Meanwhile, the propensity transformation from FCC to hexagonal close-packed (HCP) structures can be suppressed by prescribing a transformation enthalpy of $\Delta H^{fcc \rightarrow hcp}$ = 3.82 meV. The <112>-direction slips on {111} planes, as an essential mechanism for the buildup of complex deformation-induced microstructures, are primarily governed by the carved generalized planar fault (GPF) energy curve (Fig. 1F), which quantifies the energy change along the slipping path. The parameters, $\gamma_{sf}$ and $\gamma_{usf}$, largely determining the shape of the GPF curve, are sensitive to alloy composition and critical fitting parameters for the TWIP steel potential.

A number of MD simulations have been carried out previously on deformation twinning in metals or alloys with low SFE, which is known to be crucial for the remarkable high work hardening rates observed in experimental studies [33-37]. Here, we have performed MAMD simulations of fully three-dimensional polycrystals with an average grain size of 20 nm (Fig. S2). Details about the simulations are given in the Methods. At the initial stage, the plastic deformation is mainly carried by the gliding of partial dislocations nucleated from grain boundaries. The associated SFs either transect the whole grain or are stopped by other inclined SFs (Fig. 2*A*), which is consistent with the typical transmission electron microscopy (TEM) observation shown in Fig. 2*B*. Deformation twinning is more frequently observed at larger plastic strains (Fig. 2*C-E*). The primary twin structures are formed in grains with one dominant slip system being activated (Fig. 2*C*). Due to the random texture, competition among various slip systems leads to multiple deformation twin variants (Fig. 2*E*). These microstructures match well with our own TEM images on severely deformed TWIP steels (Fig. 2*D* and 2*F*) and those in the literature [35-37]. It is noted that a high density of extended dislocations also serve as important carriers of plastic deformation in the TWIP steel, as revealed by both MAMD simulations (Fig. 2*G*) and experiments (Fig. 2*H*). So far, the mechanical behaviors of low SFE alloys as determined by the dynamic interactions among these typical microstructures have been largely investigated from theoretical viewpoints and experimental evidences [35,37-39]. The fact that our MAMD simulations are fully consistent with the existing studies provides a validation of the proposed meta-atom method.

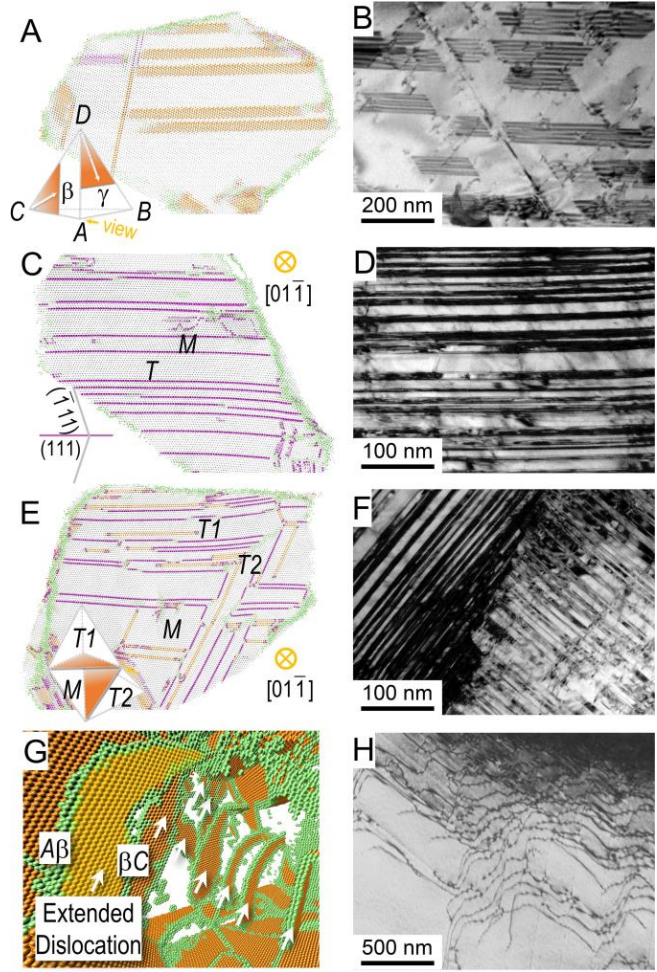

**FIG. 2.** Comparison of typical deformation-induced microstructures by MAMD simulations and TEM observation at different deformation stages. (*A-B*) Long SFs transecting a whole grain at strains less than 10%; (*B-E*) Formation of primary and secondary twin structures at medium to high strains (30% ~ 50%); (*E-F*) Abundant extended dislocations in both simulations and experiments. Insets to (*A*) and (*C*) are Thompson tetrahedrons, illustrating SF orientation and primary-secondary twin relation in FCC lattice, respectively. Color method: SF in orange; TB in purple; both partial dislocation and GB in green.

The deformation twinning process involves intensive slips of perfect and partial dislocations, as observed in both experiments and MAMD simulations. Interactions between twinning and gliding dislocations at TBs are believed to make twinning an effective strategy in simultaneously increasing both strength and ductility of TWIP steels [37]. To classify the twin-slip and twin-dislocation interactions in simulations, we adopt a double tetrahedron notation proposed by Hartley and Blachon (Fig. 3*A*) [40] and consider the following interactions.

(1) **C$\delta$**. Gliding of Shockley partials (*e.g.* *C*$\delta$) along TBs ($\delta$), the so-called twinning partials, contributes significantly to deformation twin nucleation and growth, as well as the detwinning process. Figure 3*B* shows a twin lamella formed by repetitive emission of twinning partials from GBs onto successive (111) stacking planes, according to a layer-by-layer growth mode. Other nucleation mechanisms are presented in Fig. S3. A three-layer twin embryo can be nucleated at GBs by emitting a group of three twinning partials on adjacent planes (Fig. S3*A*) [37,41], which are of the same type and remain separated. Double Shockley partials, as proposed by Hirth and Hoagland [42], can be formed by combining two partials with different Burgers vectors on neighboring glide planes (Fig. S3*B*). Occasionally, two intrinsic stacking faults (ISF) overlap and transform local FCC lattice into an unstable four-layer HCP phase (Fig. S3*C*). The inner two HCP planes, one from each of the two ISFs, are quickly transformed back to FCC planes via gliding of a twinning partial, leaving a four-layer twin lamella. This energetically driven process has been naturally incorporated in the mea-atom potential by design. It is noted that GB networks are prolific partial dislocation sources in low-SFE metals [9-11,13,43-45], as observed in our simulations. Movie S1 reveals a high mobility TB migration process assisted by a flooding of GB-nucleated twinning partials, in spite of a high-density of defects stored in the TBs. As a consequence, this mechanism dominates twin formation and TB migration, which accounts for a significant fraction of the total plastic strain.

(2) **C$\alpha$**. Intersections between SFs and TBs are frequently observed in grains where two or more slip systems are activated. Twin boundaries are effective barriers for impinging

30° Shockley partials (*e.g.* Cα), which may retract, stay at or interact with a TB according to [33,34,46,47]

$$C\alpha \rightarrow C\delta + \delta\alpha \tag{3}$$

where δα corresponds to a net stair-rod dislocation. This process is not spontaneous and only occasionally observed in our simulations (Fig. 3*C*). The twinning partial (Cδ) associated with the dissociation of a 30° Shockley partial (*e.g.* Cα) according to (1) can only glide along one of the permissible slipping directions, which are opposite for planes I and II shown in Fig. S4. Slipping on plane I (II) increases (decreases) the twin thickness by one layer and, consequently, may lead to different energy change under a given local stress state. Therefore, only certain 30° Shockley partials can react with TBs according to (1).

It is also found that a TB can retain a high migration mobility even when intersected by a high-density of SFs (Movie S1). Here, the most frequently observed dislocation reactions are

$$C\alpha + \delta C \rightarrow \delta\alpha \tag{4}$$

$$C\delta + \delta\alpha \rightarrow C\alpha \tag{5}$$

Reaction (4) gives similar observation as and is easily confused with reaction (3), especially in a two-dimensional view (Fig. S5), but unlike (3), it takes place spontaneously. A stair-rod dislocation (δα), appearing as an atomic step at the intersection, can be removed following reaction (5), which is also spontaneous. Sketches of the above reactions are shown in Fig. 3*E-F* and Fig. S5. Assisted by prolific GB dislocation sources, repeating reactions (4) and (5) in sequence leads to fast migration of TBs, as observed in Movie S1. Reaction (4) serves as an elementary mechanism in forming more complex twin structures.

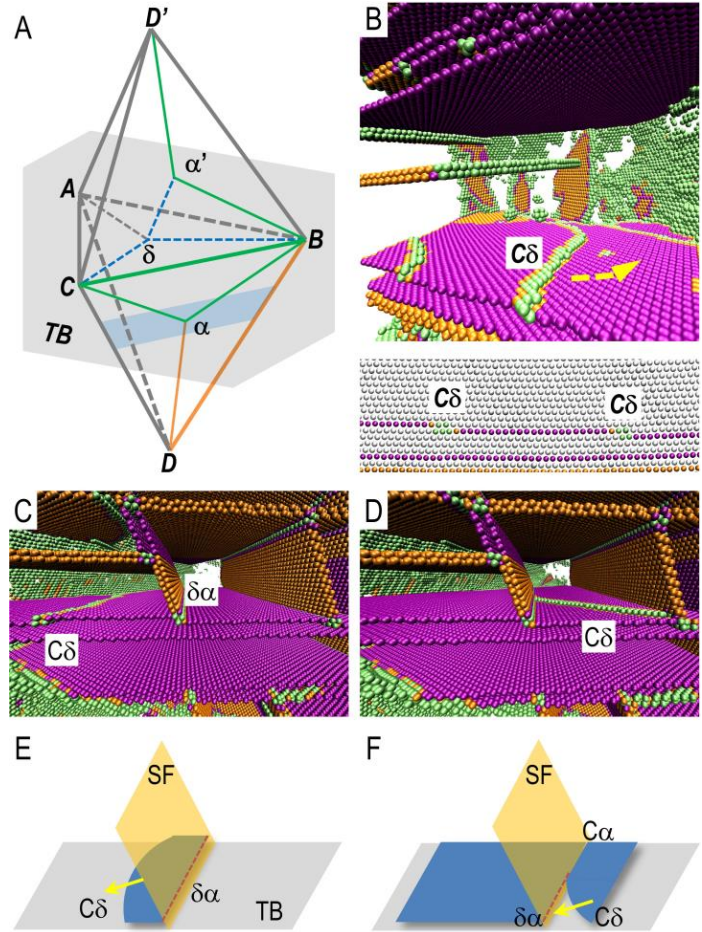

**FIG. 3.** Twin-slip interactions. (*A*) The double Thompson's tetrahedron with δ plane as the coherent TB. In derivation of the reactions in the text, all the matrix dislocations, unless specified, are considered to slip in α plane and the twinning partials glide in δ or δ' plane. (*B*) A deformation twin grows layer-by-layer with the successive emission of Cδ on adjacent slip planes from a GB. (*C*) A partial dislocation nucleates at and propagates from the intersection between SF and TB. (*D*) Sweeping of the GB-nucleated partial dislocation removes the stair-rod dislocation. (*E-F*) schematically show the corresponding processes in (C) and (D), respectively.

(3) **CB(α)**. If the requirements for reaction (2) is not satisfied, a trailing Shockley partial (αB) can be partly driven close to the leading partial (Cα), forming an extended screw dislocation segment (Cα + αB → CB(α)) (Fig. 4*A-B*). Constriction and re-dissociation onto the twinning plane facilitate a cross-slip process (CB(δ) → Cδ + δB) and create a local constricted node connecting CB(δ) and CB(α). A complete cross-slip is finished by a step-by-step fast moving of the node, significantly lowering the energy barrier. Figure 4*C* schematically shows the corresponding process. For a twin lamella with four-layer thickness, the extended dislocation CB(δ) is in contact with both TBs. Another cross-slip from δ to α' will make CB(α) completely transmitted through the twin, leaving TBs intact (Fig. 4*D-F*). Sequential cross slips are required to realize transmission in a thicker twin, which is rarely observed. For most of the time, extended dislocations are trapped inside twins, increasing the stored defect density.

(4) **DB(α)**. The 90° Shockley partials (Dα) are effectively blocked by TBs. Combining with a trailing partial dislocation (*e.g.* αB), it can transect the twin on α' plane with Burgers vector Cα' and leaves a stair-rod dislocation (δα') on the TB (Fig. 4*G-H*). The dislocation reaction can be written as

$$DB(\alpha) \rightarrow C\alpha' + \delta\alpha' \qquad (6)$$

It is noted that this reaction lowers the total energy and transforms a 90° Shockley partial (Dα) into a 30° Shockley partial (Cα'), signifying the importance of reaction (4) and (5). The corresponding 2D views are shown in Fig. 4*I-J*.

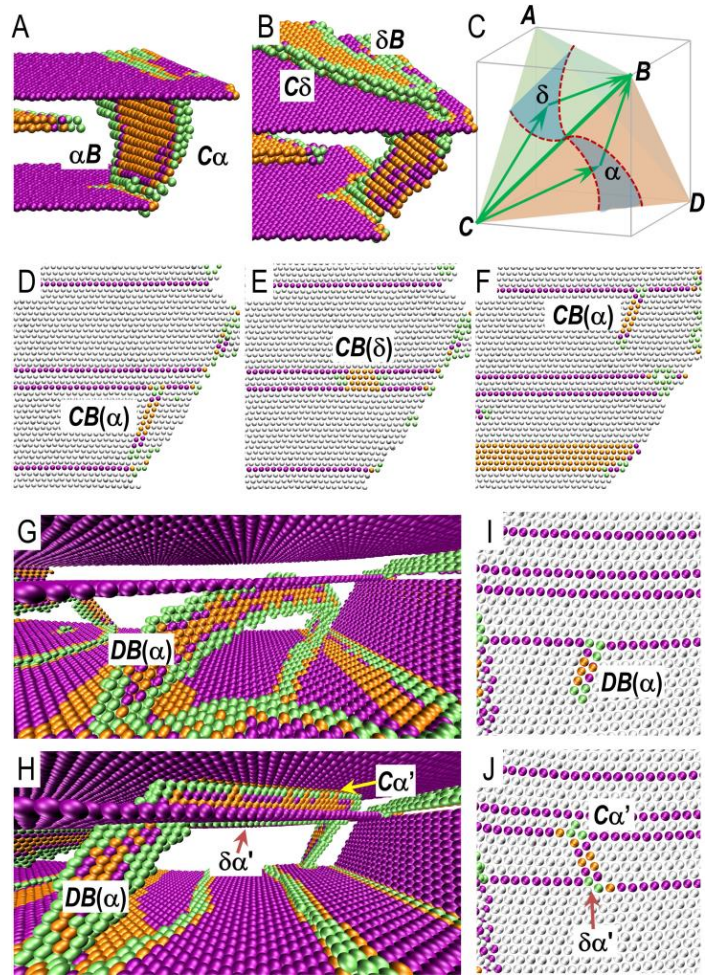

**FIG. 4.** Twin interactions with extended dislocations. (*A*) Intersection between a TB and an extended dislocation CB($\alpha$). (*B*) Cross-slip of CB($\alpha$) onto $\delta$ plane. (*C*) Representation of the cross-slip using Thompson's tetrahedron. (*D-F*) Transmission of a screw dislocation through a four-layer twin lamella in a sequence of cross-slip processes. (*G-H*) Cross slip of an extended dislocation with edge component. The corresponding 2D views are shown in (*I-J*), respectively.

Deformation twinning leads to dynamic grain refinement with TBs as effective barriers (Fig. 5*A-B*). The grain is separated into two parts by a transecting twin lamella. Multiple slanted partial dislocations impinge onto the TB, causing severe lattice distortion. The local strain is relaxed after nucleating a secondary twin embryo according to reaction (4) and (5). The V-shaped twin structure forms by nucleating a secondary twin separated from the primary twin by an array of stair-rod dislocations (Fig. 5*C-D* and Movie S2). The as-formed twin structures further partition the grain, which is expected to enhance the strength.

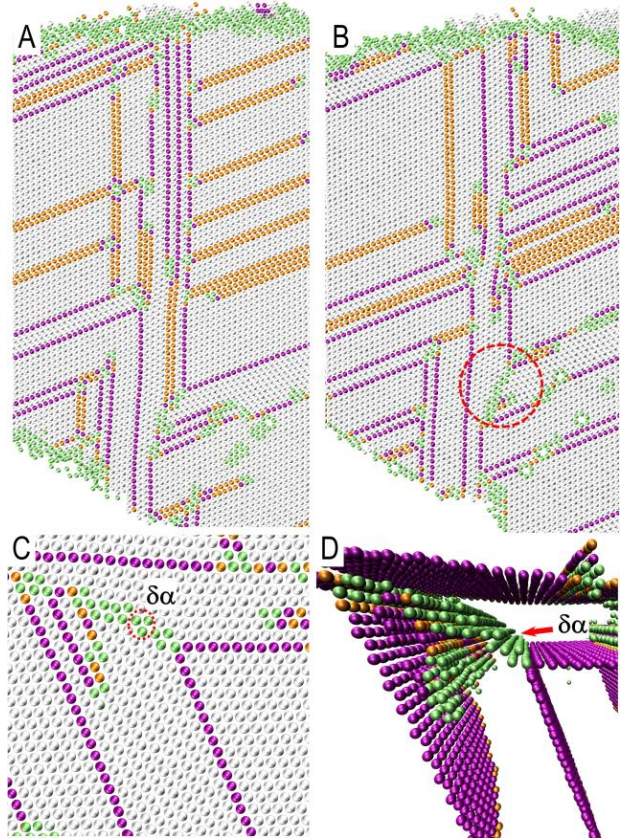

**FIG. 5.** Deformation twinning leads to partition of grains. (*A*) Intersection of SFs with TBs from both sides. (*B*) Secondary twin lamellae nucleate upon intensive slip of slanted partial dislocations. (*C*) Intersection between primary and secondary twins, as circled in (*B*), consists of an array of stair-rod dislocations. (*D*) Three-dimensional view of the stair-rod array.

A high-density of forest extended dislocations (Fig. 2*G-H* and Fig. S6) is another important feature of cold-worked TWIP steels. Constricted dislocation jogs are formed by transecting these dislocations with SFs (Fig. S6*A-B*). A super jog can be observed at the intersection between a twin lamella and an extended dislocation (Fig. S6*D-F*). Non-conservative jog motion produces vacancies (v) either individually or in the form of tubes

(Fig. S6*C* and *G*), which has been extensively studied in Cu [14,48-50]. Our simulations reveal that similar vacancy tubes are more easily formed in TWIP steels (Fig. 6), which is characterized by SF interactions especially at the early deformation stage. We propose an opposite double tetrahedron notation (Fig. 6*G*) to classify the observed SF interactions with the assumption that a SF is already present along the $\alpha$ (or $\alpha'$) plane, denoted as SF($\alpha$). The incident leading and trailing Shockley partials slip on $\delta$ (or $\delta'$) plane, marked with symbol ⌢ and ⌣, respectively. Penetration by a 90º Shockley partial, with Burgers vectors of $\widehat{A\delta}$ or $\widehat{\delta A}$, will geometrically generate a vacancy (v) or interstitial (i) tube, respectively. The associated high energy barrier makes single SF($\alpha$) an effective obstacle to slip. For a 30º Shockley partial (*e.g.* $\widehat{\delta C}$), a stair-rod dipole $\delta\alpha$-$\delta'\alpha'$ forms at the intersection, as indicated in Fig. 6*D* and *G*. It is noted that the slip vector $\delta C$ is equal to $\delta'B$ on the right side of the $\alpha$ plane in this opposite double tetrahedron setup (Fig. 6*G*). Two Frank dislocations with opposite direction are sequentially generated according to A'$\delta'$ + $\delta'\alpha'$ → A'$\alpha'$ (Fig. 6*B* and *E*) and A$\delta$ + $\delta\alpha$ → A$\alpha$ (Fig. 6*C* and *F*). A vacancy tube is formed by annihilating A$\alpha$ and A'$\alpha'$ located on neighboring planes. Such tubes may provide fast diffusion channels for small solute atoms, such as C, N, Si or Al in steels. It is expected that more pronounced bake hardening effect may take place in low-SFE alloys than high-SFE ones at the early deformation stage. The leading and trailing slips shift the SF($\alpha$) by vector AC (or A'B). Geometrically, reversing the slip vectors will create a line of interstitials, which is not observed in our simulations. If B$\delta$ is the trailing Shockley partial, the stair-rod dipole will be erased and the lattice recovered to the original state (o). The SF interactions are summarized in (7 - 9).

$$\widetilde{A\delta} + \widehat{\delta C} \xrightarrow{SF(\alpha)} v \qquad (7)$$

$$\widetilde{\delta A} + \widehat{C\delta} \xrightarrow{SF(\alpha)} i \qquad (8)$$

$$B\delta + \delta C \xrightarrow{SF(\alpha)} o \qquad (9)$$

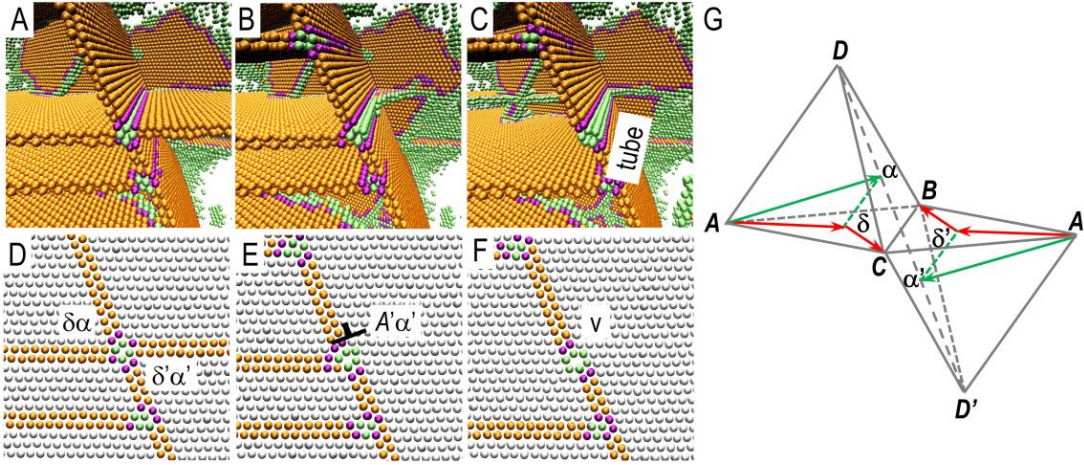

**FIG. 6.** A vacancy tube formation mechanism by SF interactions. (*A*) Intersection of two SFs and dislocation reactions (*B-C*) lead to the formation of a vacancy tube. The corresponding 2D views are shown in (*D-F*), respectively. (*G*) Opposite double tetrahedron notation.

In contrast to metals or alloys with medium to high SFEs, the deformation mechanism of TWIP steels, with SFE as low as 19 mJm$^{-2}$, is mainly characterized by SF interactions, deformation twinning and intensive extended dislocation activities. Whether the dislocation activity is dominated by extended partials or full dislocations strongly depends on the ratio $\gamma_{sf}$ to $\gamma_{usf}$, both of which are critical fitting parameters in optimizing the meta-atom potential. Similarly, the twinning propensity *via* gliding Shockley partials on adjacent planes is dependent on the ratio $\gamma_{utf}$ to $\gamma_{usf}$. A close relation between $\gamma_{utf}$ and $\gamma_{usf}$ can be derived based on a first-order approximation method:

$$\gamma_{utf} \approx \gamma_{usf} + \tfrac{1}{2}\gamma_{sf} \qquad (10)$$

which is verified by the available data from calculations based on first principle methods, tight bonding and MD simulations. Detailed discussion can be found in the online supporting material. The energy barriers are always in the order of $\gamma_{usf} > \gamma_{utf} - \gamma_{sf} > \gamma_{usf} - \gamma_{sf}$ for FCC metals in generating a single SF, a twinning partial and a trailing partial dislocation, respectively. Since slip process is thermally activated, the relative occurrence probability of different mechanisms is exponentially proportional to the difference in barrier heights ($\gamma_{sf}$ or ½ $\gamma_{sf}$). This directly leads to the conclusion that emission of a trailing partial dislocation is energetically more favorable than nucleation of a deformation twin. However, a small $\gamma_{sf}$ (*e.g.* 19 mJm$^{-2}$ for the TWIP steel under consideration) greatly promotes deformation twining, as revealed by our simulations. The complex twin-slip and twin-dislocation interactions demonstrate the dual role of deformation twins as both dislocation barrier and storage, potentially contributing to the high strength and ductility of TWIP steels.

## IV. CONCLUSION

In summary, a meta atom method has been proposed to unify MD simulations of both pure metals and alloys in the framework of EAM potentials. The universality of the EAM potential provides adequate flexibility in constructing model materials with properties varying in a sufficiently large range, covering spectra of different classes of metallic alloys. In this paper, we have developed a meta-atom potential for a TWIP steel with SFE as low as 19 mJm$^{-2}$. Fully three-dimensional MD simulations reveal that interactions between slipping and deformation twinning can account for most of the microstructural evolutions observed in experiments. We expect that the calculation of macroscopic measurable quantities for engineering alloys can be incorporated in meta-atom potentials in this way, enabling massively parallel MAMD simulations to shed light on the microscopic deformation mechanisms and facilitate the construction of materials with specific mechanical properties. This approach may pave a practical way to understand the mechanical behaviors of complex alloys taking full advantage of the rapidly increasing computational power.

**ACKNOWLEDGMENTS.** H Wang acknowledges the financial support from the Natural Science Foundation of China (Grant No. 11322219 and No. 11321202). The computational work carried out by TianHe-1(A) system at National Supercomputer Center in Tianjin is gratefully acknowledged.

# Supporting Online Material

New insights on the stacking fault behavior in twin induced plasticity from meta-atom molecular dynamics simulation

**Table of Contents**

## 1. Applicable size regime of the meta-atom method

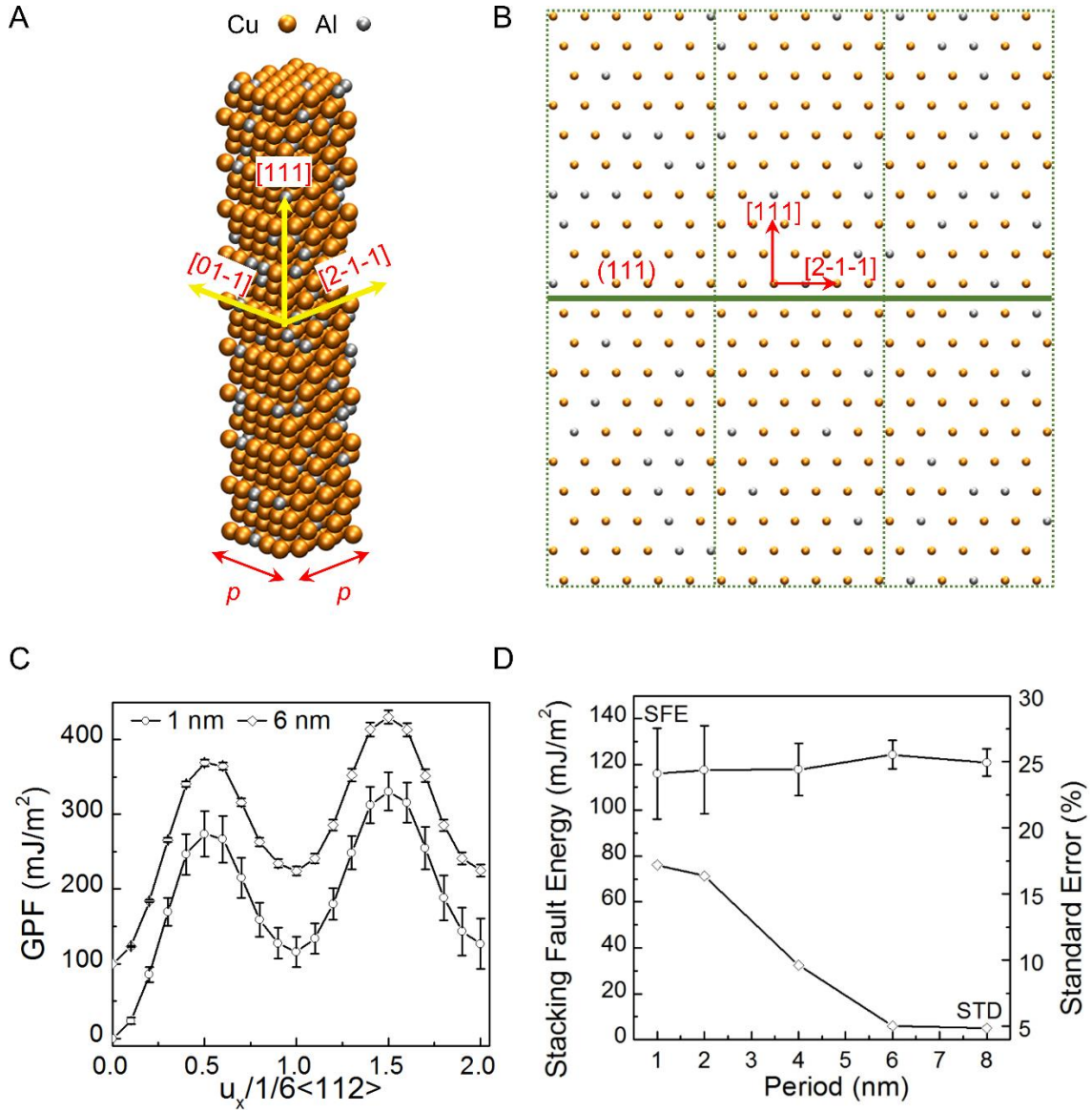

**Fig. S1.** (*A*) Atomic model for MD simulation of aluminum bronzes (Cu-6wt.% Al). The simulation box period is denoted as *p*. (*B*) The projection of the simulation system along [01-1]. The dashed line and the solid line indicate the simulation box and the slip plane, respectively. (*C*) The generalized planar fault (GPF) energy curves for simulation boxes with periods of 1 nm and 6 nm. The curves are shifted from each other for clarity. (*D*) The variation of SFE with the simulation box period.

The applicable size regime of the meta-atom method can be defined by considering the fluctuation in GPF curves due to chemical inhomogeneities in alloys. To demonstrate this, we take aluminum bronzes as the model system (Fig. S1*A-B*). The interatomic EAM potentials from Liu *et al*. (1) are used to model Cu-Cu, Cu-Al and Al-Al interactions. Figure S1*C* shows the GPF curves for the atomic models of Cu-6wt.% Al with simulation box periods (*p*) of 1 nm and 6 nm, respectively. A variation in GPF energy is observed from test to test due to the local atomic arrangements near the slip plane. It is interesting to observe that the average SFE is nearly independent of the box period, while the standard error is reduced to 10% when *p* > 4 nm and 5% when *p* > 6 nm, suggesting a critical length scale of 4 – 6 nm (Fig. S1*D*). It is noted that this length is the same for aluminum bronzes with different composition, which is slightly larger than the dislocation core size (1 – 2 nm), but much smaller than other characteristic sizes such as the average grain size and distance between dislocations.

## 2. The optimization scheme for meta-atom potentials

We followed the established method of Mendelev *et al* and Ackland *et al*. (2-5) to develop meta-atom potentials. The embedding function and the pair function are formulated by superposition of basis polynomial functions:

$$F(\rho) = a_1\sqrt{\rho} + a_2\rho^2 + a_3\rho^4 + H(\rho - 40) \cdot 0.1 \cdot (\rho - 40)^2 \tag{1}$$

$$\phi(r) = (\sum_{i=0}^{2} b_i r^i) \cdot H(r_0 - r) + H(r - r_0)\sum_{i=1}^{N}\sum_{p=3}^{4} c_{i,p}(r_i - r)^p H(r_i - r) \tag{2}$$

where $H(r)$ is the Heaviside function, $N$ the number of basic functions and $a_i$, $b_i$, $c_{i,p}$ and $r_i$ are the fitting coefficients. We used 8 knots for the pair function. The $\rho$ function is adopted from M.I. Mendelev *et al* (3), which accurately reproduces the crystalline, melting properties and liquid structure of iron. The cutoff distance is chosen as $r_{cut} = 1.65\ a_0$. The pair function and density function are forced to be 0 at a cutoff distance. The knots and coefficients for the meta-atom potential of selected TWIP steel are given in Table S1 and S2, respectively.

**Table S1.** Knots for the pair function.

| No. | $r_i$ (Å) |
|---|---|
| 0 | 2 |
| 1 | 2.4 |
| 2 | 2.8 |
| 3 | 3.2 |
| 4 | 3.5 |
| 5 | 4 |
| 6 | 4.6 |
| 7 | 5.4 |
| 8 | 6 |

**Table S2.** Summary of the fitting coefficients.

| | TWIP steel | | TWIP steel |
|---|---|---|---|
| $a_1$ | -9.500000000000000e-01 | $c_{3,4}$ | 4.293559073390630e+00 |
| $a_2$ | 1.618465111936526e-03 | $c_{4,3}$ | 3.791895483309166e+00 |
| $a_3$ | -6.762905624654250e-07 | $c_{4,4}$ | -3.670595358210705e+00 |
| $b_0$ | -6.277114387582506e+00 | $c_{5,3}$ | 1.899517741423221e-02 |
| $b_1$ | 0 | $c_{5,4}$ | -1.222018433436698e+00 |
| $b_2$ | 5.000000000000000e+01 | $c_{6,3}$ | -4.028413605605736e-01 |
| $c_{1,3}$ | 7.410332271283292e+01 | $c_{6,4}$ | 3.695323363306102e-01 |
| $c_{1,4}$ | 6.848516616595713e+02 | $c_{7,3}$ | -4.338132774276135e-02 |
| $c_{2,3}$ | -2.163946029666258e+00 | $c_{7,4}$ | 1.197693208633643e-02 |
| $c_{2,4}$ | 7.924697619521995e+00 | $c_{8,3}$ | 2.600124439135270e-02 |
| $c_{3,3}$ | 3.949239092637880e+00 | $c_{8,4}$ | -5.584234894549544e-03 |

## 3. Relationship between $\gamma_{sf}$, $\gamma_{utf}$ and $\gamma_{usf}$

In the framework of the EAM formalism, the energy of one atom is given by

$$U_i = F(\rho_i) + \frac{1}{2}\sum_{j \neq i} \phi(r_{ij}) \tag{3}$$

where $F$ is the embedding function and $\phi$ is the pair function between atoms. To first order approximation, the embedded energy variation can be neglected during slipping, and $\gamma_{sf}$, $\gamma_{utf}$ and $\gamma_{usf}$ can be calculated in the same form as

$$\gamma = (\frac{1}{2}\sum_m N_m \phi(\alpha_m \cdot a_0))/S \tag{4}$$

where subscript $m$ denotes the $m^{th}$ nearest neighbors, $N_m$ and $\alpha_m$ are the number of the $m^{th}$ nearest neighbors and the corresponding geometric factor, respectively, and $S$ is the projected area of one atom on a close-packed plane. We note that $N_m$ are different for the stacking, unstable stacking and unstable twinning configurations, while $\alpha_m$ are identical for all configurations. The values of $N_m$ and $\alpha_m$ are given in Table S3. Clearly, for arbitrary $m$,

$$N_m \text{(utf)} = N_m \text{(usf)} + \frac{1}{2} N_m\text{(sf)} \tag{5}$$

which directly leads to

$$\gamma_{utf} \approx \gamma_{usf} + \frac{1}{2}\gamma_{sf} \tag{6}$$

Surprisingly, this estimation gives good predictions of $\gamma_{utf}$, as verified by the available data based on first principle (Table S4), tight bonding (Table S5) and MD (Table S6) calculations.

**Table S3.** Geometric factors and number of $m^{th}$ nearest neighbors for calculating $\gamma_{sf}$, $\gamma_{utf}$ and $\gamma_{usf}$.

| Atomic Shell | $\alpha_m$ | $N_m$ (sf) | $N_m$ (usf) | $N_m$ (utf) |
|---|---|---|---|---|
| 1 | $\sqrt{11/24}$ | 0 | 4 | 4 |
| 2 | $\sqrt{1/2}$ | 0 | -6 | -6 |
| 3 | $\sqrt{17/24}$ | 0 | 4 | 4 |
| 4 | 1 | 0 | -6 | -6 |
| 5 | $\sqrt{29/24}$ | 0 | 8 | 8 |
| 6 | $\sqrt{4/3}$ | 4 | 0 | 2 |
| 7 | $\sqrt{11/8}$ | 0 | 4 | 4 |
| 8 | $\sqrt{35/24}$ | 0 | 4 | 4 |
| 9 | $\sqrt{3/2}$ | -12 | -24 | -30 |
| 10 | $\sqrt{13/8}$ | 0 | 8 | 8 |
| 11 | $\sqrt{11/6}$ | 24 | 0 | 12 |
| 12 | $\sqrt{15/8}$ | 0 | 8 | 8 |
| 13 | $\sqrt{47/24}$ | 0 | 8 | 8 |
| 14 | $\sqrt{2}$ | -12 | -12 | -18 |
| 15 | $\sqrt{17/8}$ | 0 | 8 | 8 |
| 16 | $\sqrt{19/8}$ | 0 | 4 | 4 |
| 17 | $\sqrt{5/2}$ | -24 | -36 | -48 |
| 18 | $\sqrt{21/8}$ | 0 | 8 | 8 |
| 19 | $\sqrt{65/24}$ | 0 | 8 | 8 |

Table S4. Calculated $\gamma_{sf}$, $\gamma_{utf}$ and $\gamma_{usf}$ by first principle

| Potential | $\gamma_{sf}$ (mJm$^{-2}$) | $\gamma_{usf}$ (mJm$^{-2}$) | $\gamma_{utf}$ (mJm$^{-2}$) | $\gamma_{usf} + \frac{1}{2}\gamma_{sf}$ (mJm$^{-2}$) | Relative error* |
|---|---|---|---|---|---|
| Ag (6) | 18 | 133 | 143 | 142 | -0.70% |
| Au (6) | 33 | 134 | 148 | 150.5 | 1.69% |
| Cu (6) | 41 | 180 | 200 | 200.5 | 0.25% |
| Ni (6) | 110 | 273 | 324 | 328 | 1.23% |
| Pd (6) | 168 | 287 | 361 | 371 | 2.77% |
| Pt (6) | 324 | 339 | 486 | 501 | 3.09% |
| Al (6) | 130 | 162 | 216 | 227 | 5.09% |
| Cu-5at.%Al (7) | 20 | 170 | 179 | 180 | 0.56% |
| Cu-8.3at.%Al (7) | 7 | 169 | 176 | 172.5 | -1.99% |

* Relative error is calculated as $[(\gamma_{usf} + \frac{1}{2}\gamma_{sf}) - \gamma_{utf}] / \gamma_{utf} \times 100\%$.

Table S5. Calculated $\gamma_{sf}$, $\gamma_{utf}$ and $\gamma_{usf}$ by tight-bonding method

| Potential | $\gamma_{sf}$ (mJm$^{-2}$) | $\gamma_{usf}$ (mJm$^{-2}$) | $\gamma_{utf}$ (mJm$^{-2}$) | $\gamma_{usf} + \frac{1}{2}\gamma_{sf}$ (mJm$^{-2}$) | Relative error |
|---|---|---|---|---|---|
| Ag (8) | 18 | 93 | 105 | 102 | -2.86% |
| Al (8) | 99 | 164 | 207 | 213.5 | 3.14% |
| Au (8) | 49 | 110 | 135 | 134.5 | -0.37% |
| Cu (8) | 64 | 200 | 236 | 232 | -1.69% |
| Ir (8) | 305 | 679 | 872 | 831.5 | -4.64% |
| Pb (8) | 30 | 98 | 108 | 113 | 4.63% |
| Pd (8) | 107 | 313 | 355 | 366.5 | 3.24% |
| Pt (8) | 270 | 388 | 521 | 523 | 0.38% |

**Table S6**. Calculated $\gamma_{sf}$, $\gamma_{utf}$ and $\gamma_{usf}$ by MD simulations

| Potential | $\gamma_{sf}$ (mJm$^{-2}$) | $\gamma_{usf}$ (mJm$^{-2}$) | $\gamma_{utf}$ (mJm$^{-2}$) | $\gamma_{usf} + 1/2\ \gamma_{sf}$ (mJm$^{-2}$) | Relative error |
|---|---|---|---|---|---|
| **Ni (9)** | 304 | 551 | 704 | 703 | -0.14% |
| **Ni (10)** | 120 | 172 | 234 | 232 | -0.85% |
| **Cu (9)** | 21 | 154 | 163 | 164.5 | 0.92% |
| **Cu (9)** | 34 | 173 | 190 | 190 | 0.00% |
| **Al (10)** | 146 | 151 | 200 | 224 | 12.00% |
| **Al (11)** | 95 | 124 | 150 | 171.5 | 14.33% |

## 4. List of supporting figures

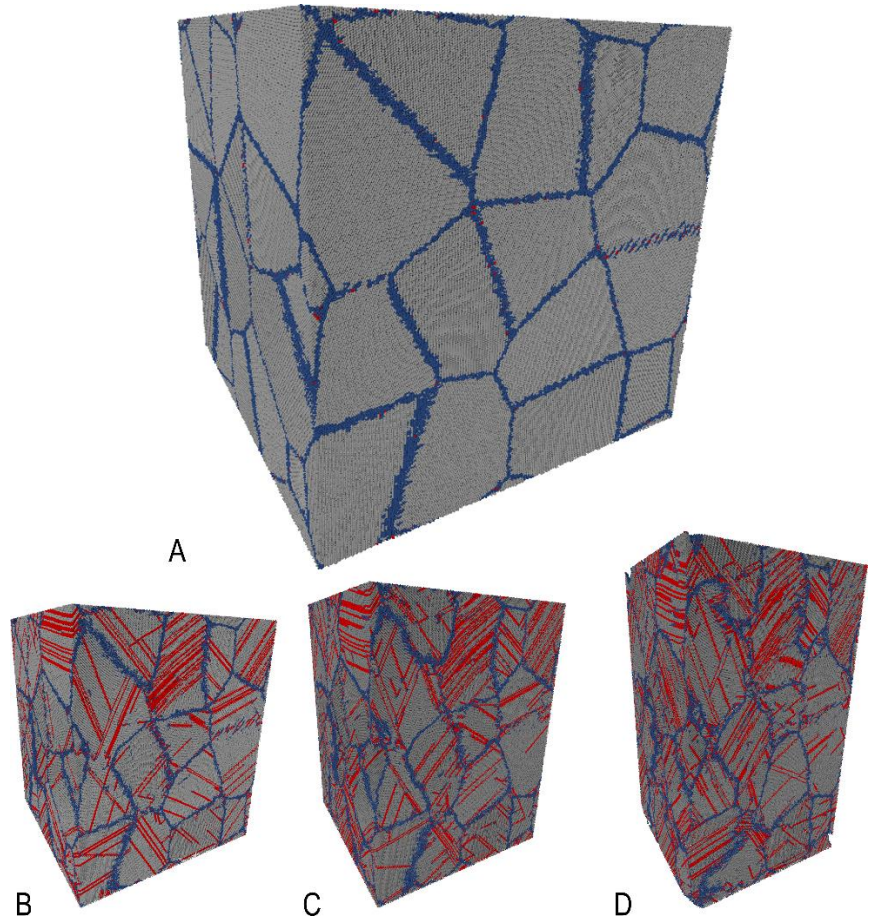

**Fig. S2.** Uniaxial deformation of a three-dimensional polycrystalline TWIP steel sample with average grain size 20 nm at different engineering strains: (*A*) 0%, (*B*) 10%, (*C*) 20% and (*D*) 40%. Color method: Gray: FCC atom; Red: HCP atom; blue: Unknown atom.

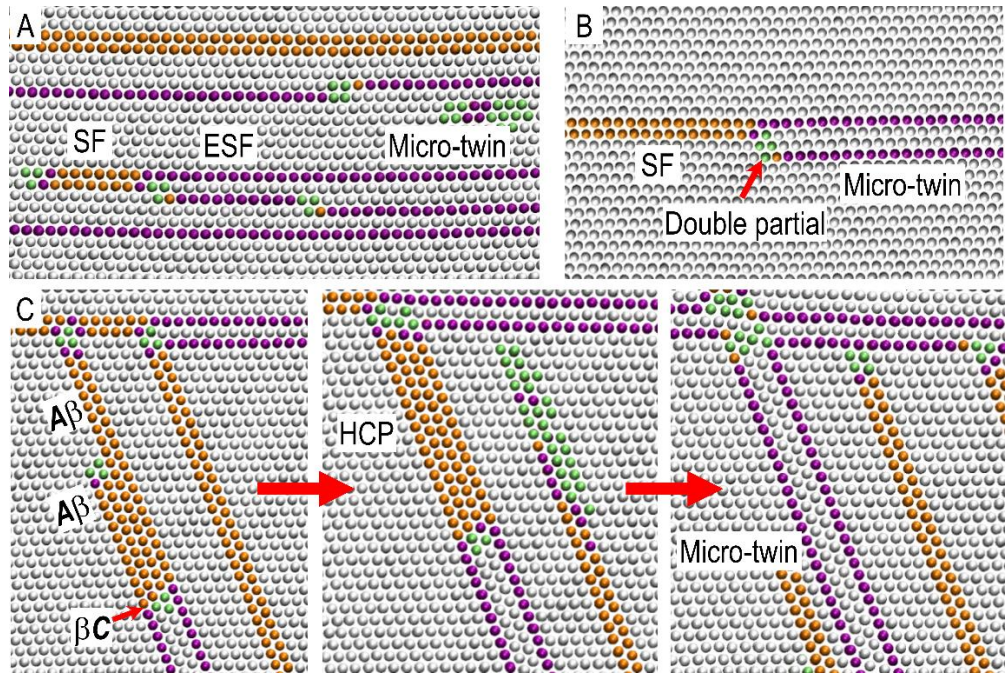

**Fig. S3.** Typical twin nucleation mechanisms. (*A*) Micro-twin is formed by successive emission of Shockley partials from the grain boundary on the right. (*B*) Double Shockley partials. (*C*) Transformation of a four-layer HCP phase into a twin lamella.

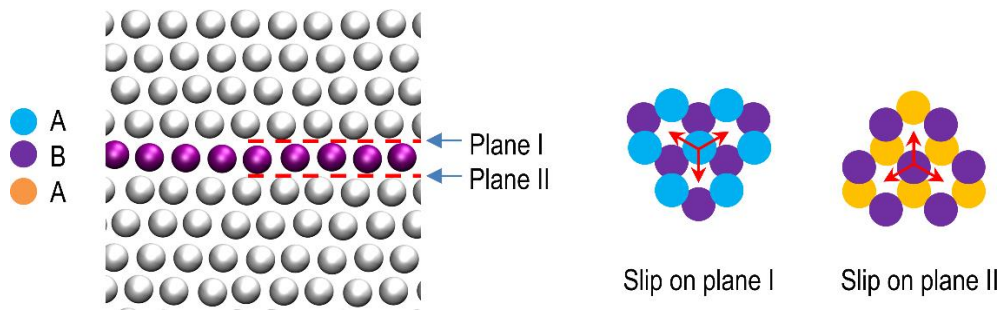

**Fig S4.** Permissible slips right above or below a TB. The slip planes are denoted as plane I and II. Clearly, the slipping directions on plane I are opposite to those on plane II.

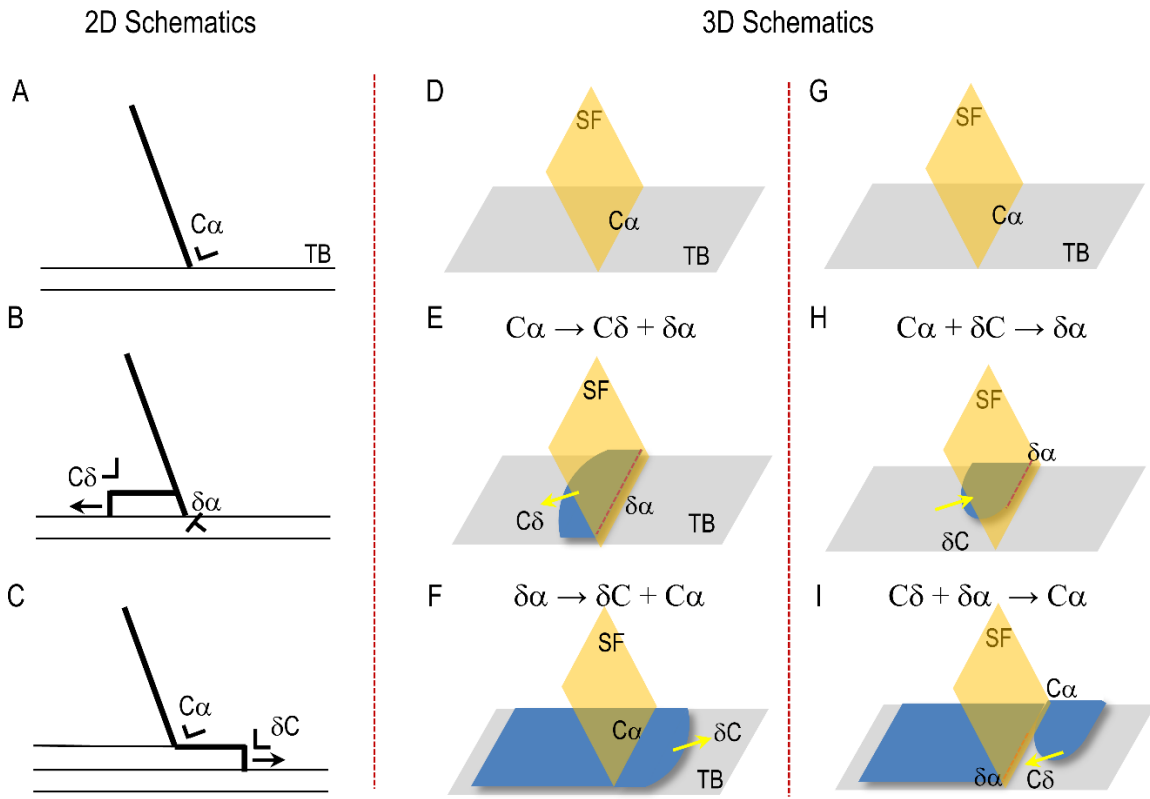

**Fig. S5.** Representations of twin-slip interactions at a TB in two-dimensional and three-dimensional views. (*A-C*) Dislocation reactions lead to migration of the TB by one layer in a two-dimensional schematic. (*D-F*) The realization of (*A-C*) by dislocation dissociation. (*G-I*) The realization of (*A-C*) by dislocation combination. It is noted that both (*D-F*) and (*G-I*) give the same two-dimensional sketch, but they correspond to different mechanisms.

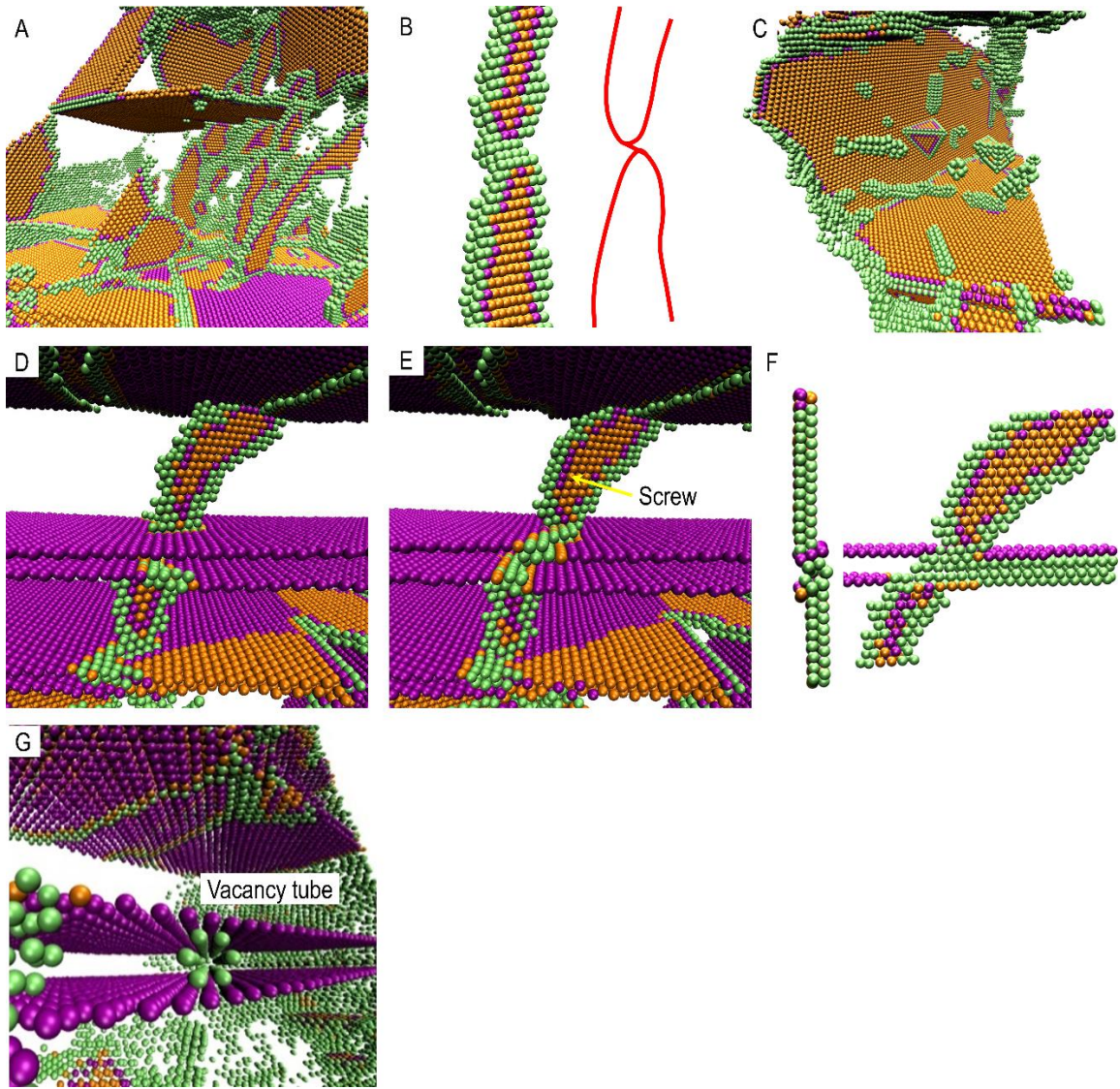

**Fig. S6.** Vacancies created by non-conservative motion of dislocation jogs. (*A*) Extended dislocation forests with jogs. (*B*) One extended dislocation with a constricted jog. (*C*) Vacancies created by non-conservative motion of dislocation jogs. (*D-E*) A super jog is formed by sectioning a screw dislocation with a twin lamella. (*F*) The side view of the super jog. (*G*) Vacancy tube is formed after the super jog moves away.

## 5. List of supporting videos

**Video S1.** Twin boundary migration process assisted by a flooding of GB-nucleated twinning partials.

**Video S2.** The V-shaped twin structure forms by nucleating a secondary twin separated from the primary twin by an array of stair-rod dislocations.